# Visualization of Band Shifting and Interlayer Coupling in $W_xMo_{1-x}S_2$ Alloys using Near-Field Broadband Absorption Microscopy

Po-Wen Tang[1], Shiue-Yuan Shiau[2], He-Chun Chou[1], Xin-Quan Zhang[3], Jia-Ru Yu[1], Chun-Te Sung[1,3], Yi-Hsien Lee[3], and Chi Chen[1*]

[1] Research Center for Applied Sciences, Academia Sinica, Taipei,115, Taiwan

[2] Physics Division, National Center for Theoretical Sciences, Taipei, 106, Taiwan

[3] Department of Materials Science and Engineering, National Tsing-Hua University, Hsinchu, 300, Taiwan

# Graphic abstract

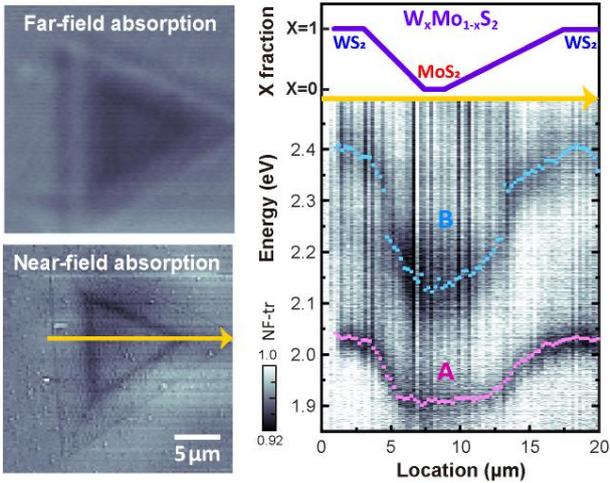

# Abstract

Beyond-diffraction-limit optical absorption spectroscopy provides profound information on the graded band structures of composition-spread and stacked two-dimensional materials, in which direct/indirect bandgap, interlayer coupling, sliding, and possible defects significantly modify their optoelectronic functionalities such as photoluminescence efficiency. We here visualize the spatially-varying band structure of monolayer and bilayer transition metal dichalcogenide alloys for the first time by using near-field broadband absorption microscopy. The near-field-spectral and -spatial diagrams manifest the excitonic band shift that results from the interplay of composition spreading and interlayer coupling. These results enable us to identify the top layer of the bilayer alloy as pure $WS_2$. We also use the aberration-free near-field transmittance images to demarcate the exact boundaries of alloyed and pure transition metal dichalcogenides. This technology can offer new insights on various layered structures in the era of "stacking science" in quest of novel quantum optoelectronic devices.

**Introduction**

Transition metal dichalcogenides (TMDs) provide an ideal two-dimensional (2D) electronic system [1-11], in which the electron spin and valley degrees of freedom can be optically or electrically controlled for optoelectronic applications. The electronic, photonic, phononic, and magnetic functionalities of these 2D semiconductors have been very much improved by the fabrication of lateral and vertical heterostructures, with possible symmetry control through the angular twisting between layers [7-15]. These tailored TMD structures give rise to spatially-varying electronic states from nano to sub-micrometer scale, hence making a highly unusual system for studying excitons, trions, biexcitons, and their many-body physics [4-10, 16-23].

So far, the characterization of direct-gap or even indirect-gap TMDs has been mostly done through photoluminescence (PL) spectroscopy [9, 13-17, 19, 23-25]. However, the PL spectroscopy can only probe the lowest exciton states, in contrast to absorption spectroscopy that can also furnish information on higher excitonic excitations [10, 23-27]. As complementary methods to PL, transmittance (or absorption) [10, 28] and reflectance spectra [25, 26, 28-31] can map the entire energy band of TMDs, from which the A, B, C excitons and the band structure can be inferred.

However, in the case of multilayers, vertically-stacked heterostructures, alloys, and highly-defective samples, the PL detection becomes very inefficient because non-radiative processes then dominate the exciton recombination. To make things more complicated, in inhomogeneous structures and alloys, the band structure changes according to the fractional composition [14, 19-23, 32] and the local coupling between layers. To analyze such composition- and location-dependent electronic states requires high-spatial-resolution spectroscopy. Yet, a severe chromatic aberration caused by broadband (or white light) illumination in far-field confocal microscopy [33] is a major hurdle to achieve high lateral absorption resolution, especially when multilayered or stacked areas of the sample are a few μm in size.

To overcome the above aberration problem in absorption mapping [34], the aperture-type scanning near-field optical microscopy (*a*-SNOM) provides an aberration-free, sub-diffraction-limit light spot. The *a*-SNOM has so far been used to optically visualize point defects, grain boundaries, and lateral heterostructures of TMDs through near-field PL mapping [35-38]. We here show that near-field absorption spectroscopy, equipped with sub-diffraction-limit spatial resolution, can provide fully-spectroscopic excitonic information, which has never been realized for TMDs up to now.

In this study, we use near-field broadband absorption microscopy to visualize the spatially-varying band shifting and interlayer coupling of monolayer TMD alloy, $W_xMo_{1-x}S_2$. The spatial distribution of its excitonic bands is reconstructed from the mapping of near-field transmittance (NF-tr). The energy-location diagrams, which reflect the electronic structure of TMD alloy, reveal the energy shifting of A, B, and C excitons along a lateral spatial profile. In the case of bilayer $W_xMo_{1-x}S_2$ alloy, the energy-location diagrams confirm that the top layer is pure $WS_2$, and the excitonic band shifting is induced by the interlayer coupling. In addition, we show that NF-tr images at different photon energies can demarcate pure TMD from alloyed with clear-cut boundaries. These results demonstrate that, to investigate the stacking and interlayer coupling of 2D materials, near-field absorption microscopy provides a powerful tool to produce aberration-free and nanoscale-resolution imaging of the band structure over a wide range of energy, even in the presence of highly lateral inhomogeneity.

**Results and Discussion**

We show in Figs. 1a and 1a' the optical pathways of far-field transmittance (FF-tr) and near-field transmittance (NF-tr) measurements. For FF-tr, the top objective lens focuses the white light source, while the bottom objective lens collects the transmitted light. However, the chromatic aberration prevents the lens from focusing the broadband light into the same spot; so, a clear contrast can only be achieved within a narrow wavelength range. In the case of NF-tr, the *a*-SNOM tip limits the amount and lateral size of light though the aperture hole, as shown in Fig. 1a', with the transmitted signal collected from far field. In our NF-tr experiments, the diameter of the aperture hole is ~150 nm.

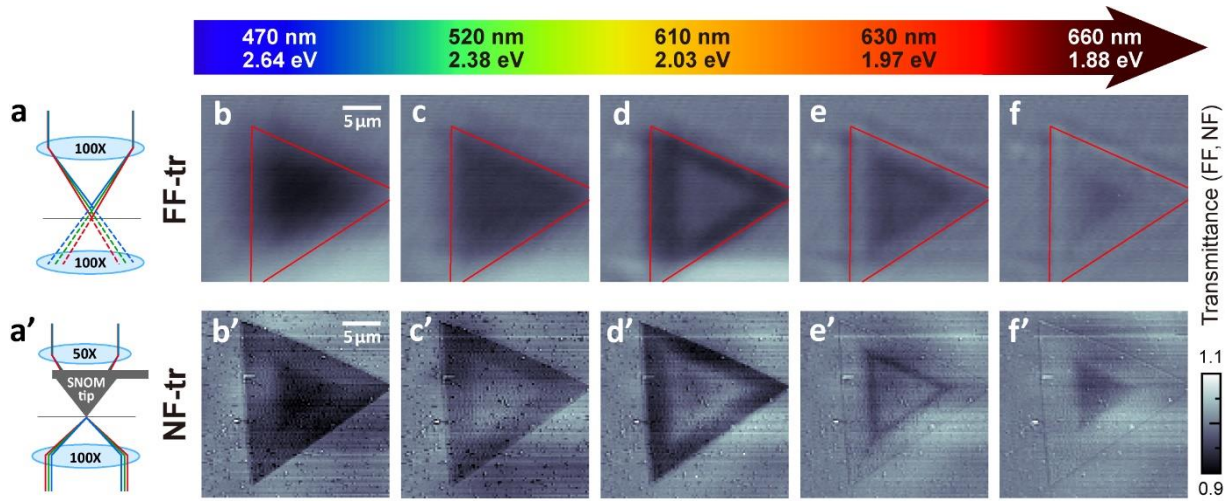

**Figure 1. (a), (a')** Optical pathways of FF-tr and NF-tr measurements. **(b-f)** FF-tr and **(b'-f')** NF-tr images of $W_xMo_{1-x}S_2$ at different photon energies. Each frame integrates a 10 nm bandwidth signal. The red triangle in (b-f) marks the location of the sample.

The $W_xMo_{1-x}S_2$ alloy was synthesized by tuning the growth temperature and reactant supplement in one-pot CVD growth [39]. Pure $MoS_2$ is located at the center of the triangular nanosheet, with the fraction of W atoms (denoted as $x$) gradually increasing from $x = 0$ at the center to $x = 1$ at the outer edges. The band shift for $W_xMo_{1-x}S_2$ can be deduced from a simple type-II band alignment between pure $MoS_2$ and pure $WS_2$, with gradient band transition in between [40]. The bandgap varies as a function of the compositional fraction, $x$, exhibiting a location-dependent spectral shift. Continuous PL peak shifting with compositional fraction has been observed by confocal PL spectroscopy of TMD alloys [14, 18-22, 32].

Fig 1. compares the FF-tr and NF-tr transmittance images at $h\nu$ = 2.64, 2.38, 2.03, 1.97, and 1.88 eV. The value of transmittance is normalized to the substrate at each wavelength. The full set of images from $h\nu$ = 2.64 to 1.85 eV (470 to 670 nm) can be found in the supplemental videos S1. We observe in FF-tr images (Figs. 1b-1f) severe chromatic aberration across a wide wavelength range: the images taken at photon energies $h\nu$ = 2.64 and 2.38 eV are highly defocused, with the triangle drifting ~1.5 μm laterally. Both phenomena are attributed to the lateral and longitudinal chromatic aberrations of the objective lens and the alignment of the broadband light path. By decreasing the photon energy, regions of low transmittance gradually shift toward the center, as shown in Figs. 1d to 1f. The best far-field focus occurs around $h\nu$ =1.97 eV, since the boundary of the triangle is the sharpest.

In contrast to FF-tr, NF-tr provides high-contrast, unblurred images (Figs. 1b'-1f') of the dark absorption regions. As the aperture on the tip determines the size and location, the NF light spot is aberration-free and highly localized, therefore making NF-tr measurement free of the aberration and resolution problems. As examples, the edge resolution in Fig. 1b' reaches ~ 300 nm, while the width of the absorption band shown in Fig. 1e' is as narrow as 1 μm. The spatial variance of the NF-tr images shown in Figs. 1b'-1f' results from the spectral shifting induced by the compositional fraction as well as the excitonic band shift.

The correct interpretation of NF-tr images, however, requires a series of NF-tr spectra. Fig. 2b shows NF-tr spectra taken across the narrow dark stripe (orange arrow) in Fig. 2a (same as Fig. 1e'). The A exciton induces a transmittance dip of 3-6%: for pure $WS_2$, the A exciton band is located at 2.03 eV, but it redshifts to 1.90 eV for pure $MoS_2$. As for the B exciton, the absorption dip that it induces is only 2-3%, with its band shifting from 2.40 eV for pure $WS_2$ to 2.12 eV for pure $MoS_2$.

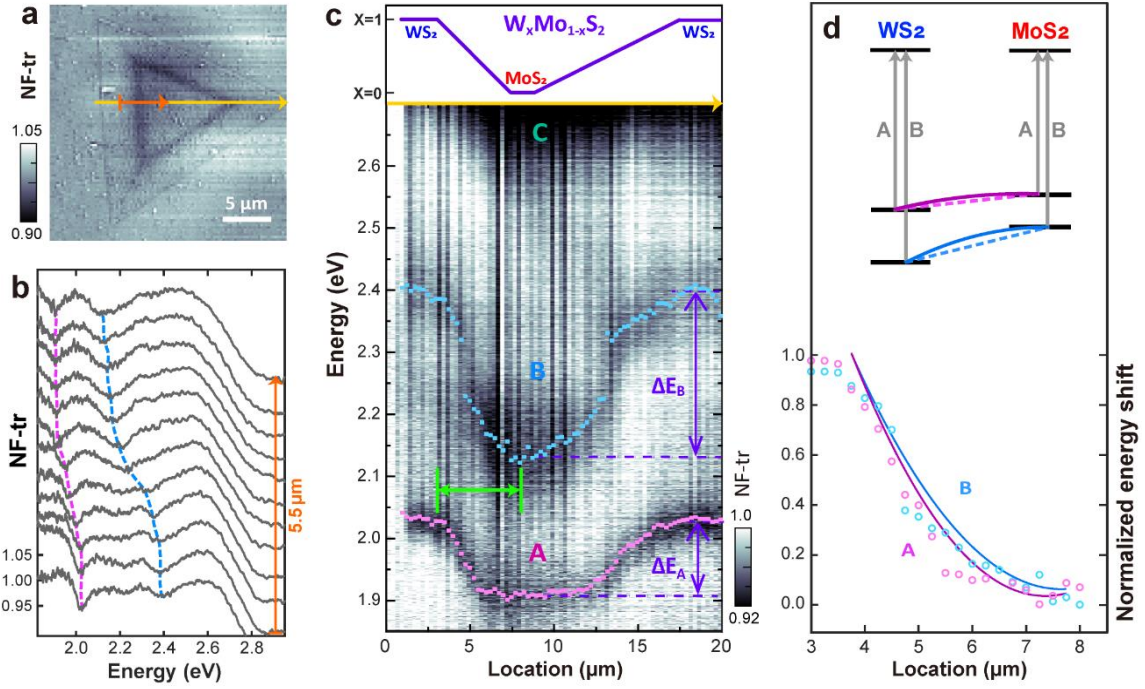

**Figure 2. (a)** The NF-tr image taken at $h\nu = 2.03$ eV (same as Fig. 1e'). **(b)** Stacking of 12 NF-tr spectra along the orange line in (a). **(c)** Energy-location diagram along the yellow line in (a). The $x$ fraction evolution with location is schematized at the top. **(d)** Normalized energy shift as a function of the location from WS$_2$ ($x=1$) to MoS$_2$ ($x=0$) in the green line region (3 to 8 μm) in (c). Top inset shows the model of the bowing factors for A and B excitons.

Fig. 2c shows the energy-location diagram taken across the center of the triangular alloy along the yellow line in Fig. 2a. This enables the reconstruction of the entire energy band of interest for the TMD alloy, that covers not only the A and B excitons, but also the much higher C exciton. The C exciton of MoS$_2$ is responsible for the significant transmittance dip when $h\nu > 2.6$ eV. All A, B, and C excitons redshift from the edge (W-rich) to the center (Mo-rich), and then shift back to the other edge. At the top of Fig. 2c, the asymmetric dependence of $x$ on the location is simply due to the orientation of the cross-sectional line along the triangular alloy.

During alloy growth, the $x$ fraction value typically varies along the growth direction (location). Yet, in most circumstances, this value tends to depend linearly on the location. The $x$ value is commonly extracted from the A exciton band-shift in the PL spectra [19-23]. However, the determination of the pure-alloy MoS$_2$ boundary ($x = 0$) is not easy. As shown in Fig. 2c, the A exciton band-shift would lead us to mark the region of pure MoS$_2$ ($x = 0$) from location L = 5.5 to 11.5 μm. We however note that the $x = 0$ region provided by the B exciton band-shift is significantly narrower by comparison, with the discernable flat-band region only ranging from L = 7 to 8 μm. The main difficulty in precisely tracking the exciton band evolution lies in the fact that the dependence of the composition fraction on location may not be linear. The difference of the flat-band regions obtained from the A and B excitons certainly renders the value of $x$ extracted from the PL spectra questionable.

From $x = 1$ (WS$_2$) to $x = 0$ (MoS$_2$), the A (B) exciton undergoes a 130 (280) meV redshift, denoted as $\Delta E_A$ ($\Delta E_B$), which results from the band alignment of WS$_2$ and MoS$_2$ (see Fig. 2c). Since this redshift is not linear in the energy-location diagram, a bowing factor, $b$, is introduced to evaluate the deviation from linearity, namely the curvature [14, 21, 32, 41, 42]. The exciton energy curves are fitted to the following quadratic equation

$$E(x) = E_{WS_2} x + E_{MoS_2}(1-x) - bx(1-x) \qquad (1)$$

where $E(x)$ denotes the exciton energy for a specific $x$ value, $E_{WS_2}$ and $E_{MoS_2}$ denote the exciton energies of WS$_2$ and MoS$_2$. So, $E(x=1) = E_{WS_2}$ and $E(x=0) = E_{MoS_2}$. The dependence of the bowing factor is illustrated in the inset of Fig. 2d. The fitted bowing factor for the A and B excitons is equal to 0.15 and 0.28, respectively, showing a much larger curvature for the B exciton since $\Delta E_B$ is twice larger than $\Delta E_A$.

A more precise comparison is to normalize the A and B exciton energies by $\Delta E_A$ and $\Delta E_B$. The resulting normalized bowing factors, $b(A)/\Delta E_A = 1.19$ and $b(B)/\Delta E_B = 0.99$, indicate instead a more curvature bending associated with the A exciton (Fig. 2d). For smaller $x$ (W<Mo), the A exciton redshifts more slowly. In this region, it becomes questionable to use the bowing factor to correlate the band shift with the $x$ value, making it difficult to determine the boundary of pure MoS$_2$.

To go further, we calibrate the nonlinearity between $x$ and location ($d$) by introducing a power $n$ according to

$$x = (d/d_0)^n \tag{2}$$

where $d_0$ stands for the distance between pure WS$_2$ and pure MoS$_2$. The fitted *n* value is close to 1 (1.02 for the A exciton and 1.00 for the B exciton), hence proving that *x* is linear with the location, as expected.

Having understood the NF-tr energy-location diagram of Fig. 2c, it now becomes straightforward to interpret the NF-tr images given in Figs. 1b'-1f'. For photon energy $h\nu$ = 2.64 eV (Fig. 1b'), the central MoS$_2$ part exhibits a particularly low transmittance dip due to the C exciton of MoS$_2$. At $h\nu$ = 2.38 eV (Fig. 1c') and 2.03 eV (Fig. 1d'), the NF-tr images highlight the WS$_2$ part at the outer edges due to the B and A excitons of WS$_2$. At $h\nu$ =1.97eV (Fig. 1e'), only a narrow dark absorption stripe appears at a particular *x* fraction of W$_x$Mo$_{1-x}$S$_2$. Finally, at $h\nu$ =1.88eV (Fig. 1f'), pure MoS$_2$ hosts such a low-lying A exciton that corresponds to the dark triangle in the center. All this demonstrates that the NF-tr spectra and images are powerful tools to resolve the spatial variance of the complicated band structure of monolayer TMD alloys.

After considering the monolayer TMD alloy, we now extend the methodology to the more complicated case of bilayer TMD alloy. Here again, NF-tr and near-field photoluminescence (NF-PL) techniques prove valuable in determining the layer number and the interlayer coupling. Fig. 3a shows the optical microscope (OM) image of a bilayer TMD alloy. The bottom (1st-L) is a 20 μm-wide alloy similar to the monolayer sample shown in Fig. 2. As illustrated in Fig. 3b, we can see a 7 μm-wide second layer (2nd-L) TMD at the top of the alloy, the composition of which, however, requires further analysis. A much smaller triangular multilayer TMD flake that lies at the center of the 2nd-L TMD shows a strong absorption even under the OM.

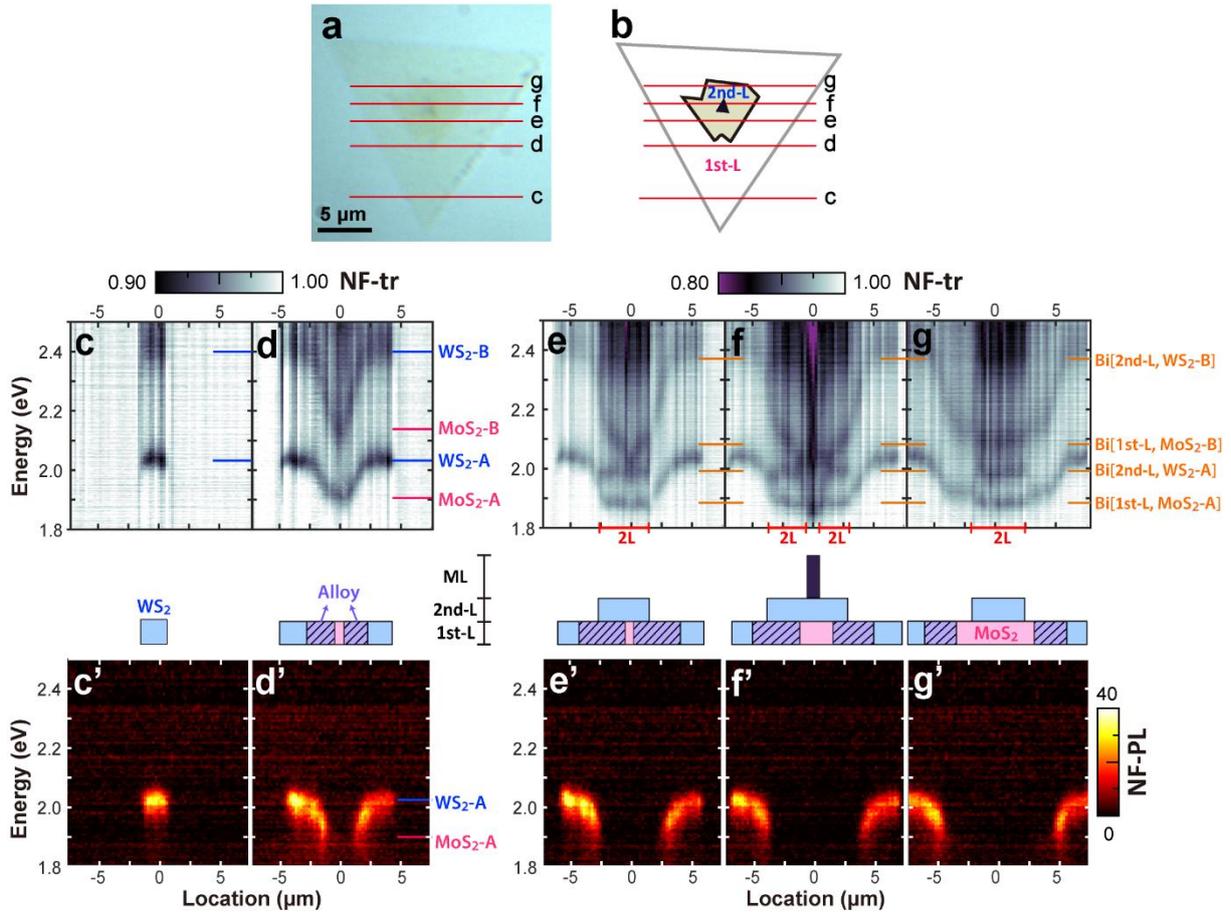

**Figure 3. (a)** OM image for multilayered alloy. **(b)** Positions of the cross-sectional lines for (c-g) and (c'-g'). **(c-g)** NF-tr energy-location diagrams and **(c'-g')** NF-PL spectral contour maps along the various lines in (b). The corresponding layer stackings for (c-g) are shown right below them. Blue and red lines stand for the excitons coming from 1st-L $WS_2$ and 1st-L $MoS_2$, respectively. Orange lines indicate the shifted bands due to interlayer coupling.

Figs. 3c-3g show the NF-tr energy-location diagrams taken along the horizontal lines (Fig. 3a) that cut across the entire alloy composition and the coupling between layers; the corresponding vertical layer stackings are shown right below. Along the same lines, the NF-PL energy-location diagrams shown in Figs. 3c'-3g' provide only limited information on the lowest electronic transition (A exciton). We now discuss the band information revealed for each line.

(i) Line-c only cuts through the pure $WS_2$ region in the 1st-L. In Fig. 3c, NF-tr reveals two flat absorption bands at 2.04 eV and 2.4 eV, corresponding to the A and B excitons of 1st-L

WS$_2$. In Fig. 3c', NF-PL only shows the emission band coming from the A exciton of WS$_2$.

(ii) Line-d cuts through the outer WS$_2$, the W$_x$Mo$_{1-x}$S$_2$ alloy in between, and the inner MoS$_2$ in the 1st-L, similar to the monolayer case shown in Fig. 2. Both NF-tr and NF-PL energy-location diagrams can resolve the band shift evolution as a function of the compositional fraction $x$. Yet, NF-PL only displays the evolution of the A exciton peak. Starting from the WS$_2$ region, the PL peak continues to redshift to 1.96 eV, as the $x$ value decreases. No PL signal comes from the central MoS$_2$-only and Mo-rich alloy regions. This leads us to suspect that the CVD growth condition is not favored for MoS$_2$, and consequently produces defective MoS$_2$. By increasing the temperature for the outer WS$_2$, a small amount of MoS$_2$ gets desorbed. Such defective MoS$_2$ results in a very low quantum yield in NF-PL. Nevertheless, even in defective areas, we can still observe the entire band shift evolution in the MoS$_2$ and Mo-rich alloy regions by near-field absorption spectroscopy, which highlights the advantage of the NF-tr (NF-abs) method.

(iii) Line-e that also cuts through the 2nd-L produces a more complicated NF-tr energy-location diagram (Fig. 3e). The one-layer (1L) region is similar to that of line-d. However, once in the 2L region, the coupling between the 1st-L and 2nd-L redshifts the A and B exciton bands of MoS$_2$ by 30 meV and 40 meV, respectively. In the 2L region, one extra flat band appears at 1.98 eV, indicating, by the flatness of the band, that the 2nd-L TMD is a pure compound, instead of an alloy. Together with the other flat band at 2.38 eV as the WS$_2$-B exciton, we can identify the 2nd-L as pure WS$_2$. The WS$_2$-B exciton is redshifted by the interlayer coupling with the bottom 1st-L, hence denoted as Bi[2nd-L, WS$_2$-B] [25, 28, 31, 43].

In NF-PL, the energy-location diagrams of Fig. 3e' and Fig. 3d' are similar, but they differ in the length of the no-emission region in the middle. Even though the 2nd-L is pure WS$_2$, the interlayer coupling with the bottom alloy layer significantly suppresses its quantum yield to become indiscernible in NF-PL. [24, 25]

(iv) Line-f cuts through the small multilayer spot in the middle, which exhibits a very strong absorption (Fig. 3f). The rest of Fig. 3f is similar to Fig. 3e, except for the bending of the Bi[2nd-L, WS$_2$-A] peak at the center. This bending is attributed to the additional coupling of the 2nd-L WS$_2$ to the multilayer flake on the top.

(v) Line-g cuts through the top edge of the 2L area. The fact that the top layer is pure WS$_2$, and the bottom layer is pure MoS$_2$ can be explained by the flatness of all four redshifted bands. This observation will be further substantiated by the NF-tr image of MoS2-B in Fig. 4c. This structure can then be seen as a WS$_2$ on MoS$_2$ vertical heterostructure. Because of the interlayer coupling, the A and B excitons of both materials undergo a significant redshift of 20-50 meV. The values of the shifted energies are: $\Delta E_{A,WS2}$=50, $\Delta E_{B,WS2}$=20, $\Delta E_{A,MoS2}$=30, and $\Delta E_{B,MoS2}$=40 meV.

While the shifting of the exciton energy generally depends on how layers are stacked [44-46], the fact that all A and B excitons in the WS$_2$/MoS$_2$ heterostructure undergo a redshift, as here observed, is somewhat puzzling at first: since the exciton binding scales inversely with the square of the dielectric constant, additional screening due to a second layer should result in a blueshift, instead of a redshift. We then note that the interlayer coupling, in conjunction with the strong spin-orbit coupling in each layer, not only can change the band gap and the band splitting, but also can modify the conduction-electron and valence-hole effective masses, which also enter the exciton binding. The observed redshift for all excitons indicates that these two effects on the conduction and valence band states can overcome the loss of binding due to the additional screening.

The schematics right below Figs. 3c-3g show the vertical layer stackings that correspond to these different lines. However, to delineate the exact boundary between pure TMD and alloy requires further information extracted from the NF-tr images at different photon energies given in Fig. 4, as we now show.

In Fig. 4a (same as Fig. 3e), we mark the energy bands associated with various excitons with the same colors used in Fig. 3. The NF-tr images taken at different photon energies are shown in Figs. 4b-4f. The scan of the full energy range is shown in the supplemental videos S2. The corresponding NF-PL images for three emission energies are shown in Figs. 4d'-4f'. All energy frames feature a strong absorption by the multilayer flake at the triangle center.

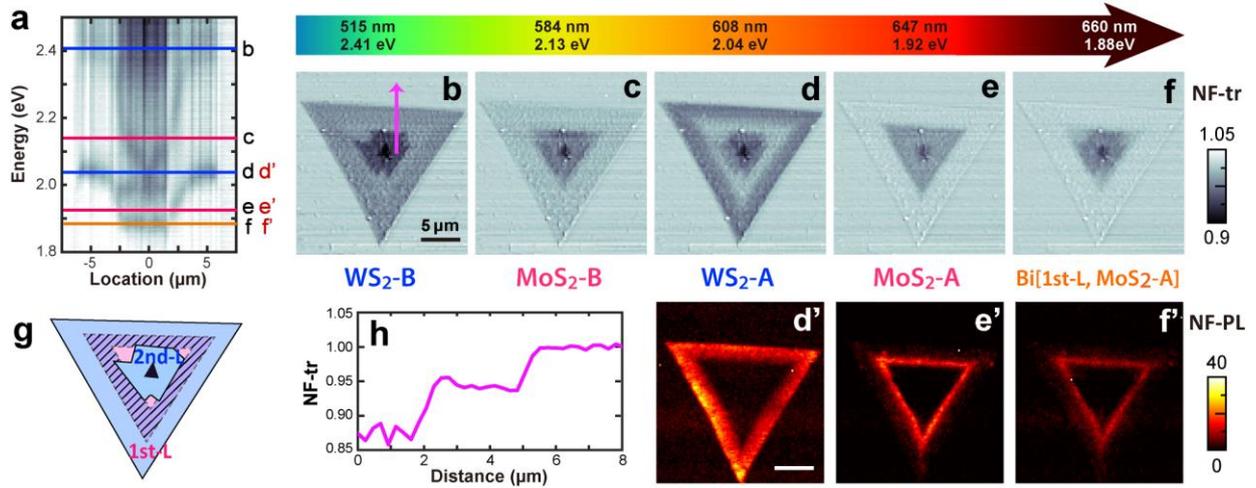

**Figure 4. (a)** NF-PL energy-location diagram for the line-e (same as Fig. 3e). **(b-f)** NF-tr images at different photon energies. **(d'-f')** NF-PL images of the A exciton at different photon energies. **(g)** Schematic of the multilayer alloy from the top view. **(h)** NF-tr profile along the pink line in (b). Each frame integrates a 3 nm bandwidth signal.

Fig. 4b shows the NF-tr image taken at 2.41 eV, which is right at the absorption energy of the $WS_2$-B exciton. Apart from the central multilayer flake, strong absorption occurs homogeneously in the 2L region. In Fig 4d, we see that the $WS_2$-A exciton not only appears in the 1st-L outer area but also in the 2L region. This is yet another evidence that the 2nd-L is pure $WS_2$. The NF-tr profile along the green line marked in Fig. 4b is shown in Fig. 4h, which manifests a 6% absorption drop for each layer at $h\nu$=2.41 eV.

The identification of pure $MoS_2$ in the alloy region is more challenging. Both the absorption of $MoS_2$-A and $MoS_2$-B excitons cover the whole 2L region (Fig. 4c and Fig. 4e), with a comparatively larger area for $MoS_2$-A. We also note from Fig. 2 that the larger bowing factor for the A exciton makes it less sensitive to the band shifting for a small $x$ value. So, the absorption image of the A exciton is not suitable for distinguishing the edges of the alloy region. Indeed, the absorption image of the B exciton, shown in Fig. 4c, displays the actual size of pure $MoS_2$ more clearly than Fig. 4e. Moreover, the redshifted band associated with the 1st-L $MoS_2$-A exciton, shown in Fig. 4f, evidences that the coupled area is precisely the same as the shape of the 2nd-L $WS_2$. The light shade at the three triangular corners comes from the $MoS_2$-A band tails of the uncoupled 1st-L.

In Figs. 4d'-4f', NF-PL is prominent in the 1st-L $WS_2$ and 1st-L W-rich alloy regions. But because of the defective $MoS_2$, the NF-PL evolution stops at some $x$ fraction; so, NF-PL images are much less informative than NF-tr images. Note that the PL distribution is not homogeneous since the PL signal is susceptible to local features, such as defects or strains [15,17]. For example, in Fig. 4d', close to the inner edge of the emission pattern, appear many dark spots that mark the locations of the defective $MoS_2$ in the alloy. In the middle, both the defective $MoS_2$ and interlayer coupling lead to an extremely low quantum yield.

With all the above information, we can understand the distribution of composition in the bilayer TMD alloy, as illustrated in Fig. 4g. The boundaries between pure $WS_2$, the $W_xMo_{1-x}S_2$ alloy, and pure $MoS_2$ in the 1st-L, and pure $WS_2$ in the 2nd-L can be determined by combining the Nf-tr energy-location diagrams (Fig. 3) and NF-tr images (Fig. 4). Indeed, NF-tr spectral mapping is a powerful technique to study spatially-variant band structures of TMDs and multilayers. It is particularly useful when PL is not available or when the information of the B-exciton is essential.

In conclusion, we used near-field transmittance microscopy to resolve the spatial profile of the complicated $W_xMo_{1-x}S_2$ bands associated with the various excitons of $MoS_2$ and $WS_2$. We have shown that this technique can overcome the problems of chromatic aberration and diffraction limit that plague far-field broadband microscopy. Whereas NF-PL can only measure the A exciton absorption and some monolayer features, NF-tr can moreover detect the B and C exciton absorption features and the composition of multilayer TMDs. This novel technology is here shown to be indispensable for characterizing the lateral distribution and the vertical combination of monolayer and bilayer TMDs.

**Methods**

**Sample Preparation.**

The $W_xMo_{1-x}S_2$ was synthesized by a one-step CVD process. A few-nanometer $MoO_3$ film was first deposited on the sapphire substrate, and $WO_3$ powder was put in a home-built quartz reactor. Both of them were then put in the tube furnace and heated to 900°C. We changed the relative position of the quartz reactor (evaporated $WO_3$) and the substrate to optimize the temperature and reactant supplement for TMD alloys growing. In this way, we could synthesize gradually alloyed interface in the $W_xMo_{1-x}S_2$ monolayer. The details of the growth can be found in [39].

**NF and FF optical measurements**

The NF-tr and NF-PL measurement was performed on the home-built SNOM instrument [47]. A xenon lamp (EQ-99; Energetiq Technology) was used as a white light source with color glass FGS600 to filter the white light spectrum. After fiber coupling, 1.7 mW of broadband light was focused by the first objective lens (50× NA 0.5) into the SNOM tip. Transmitted light was collected by the second objective lens (100× NA 0.9) and guided through a 200-μm fiber into the spectrometer (Kymera-328i, Andor) with a cooling camera (DU420A-BEX2-DD, Andor).

The *a*-SNOM tips were made of $SiO_2$ hollow-pyramid tips (Nanosensors), and the fabrication processes could be found in [47]. NF-PL was acquired in the same experimental configuration as NF-tr. 6.5 mW of 532 nm DPSS laser was focused into the tip, and emitted PL was accumulated with typically 0.2 second integration time at each pixel. FF-tr was performed under 0.2 mW of broadband illumination on the same homemade SNOM microscope without the SNOM tips.

**Calculation of transmittance.**

For NF-tr and FF-tr measurements, the transmittance is defined as $T(x, y) = I_s(x, y)/I_{ref}(y)$, where $(x, y)$ are the coordinates for mapping, $I_s$ is the sample signal, $I_{ref}$ is the reference signal, which is obtained by averaging the first five data points at the substrate region for each mapping line. In this way, the long-term drift of transmittance can be eliminated. The spectra in Fig. 2c were smoothed with 50 pixels (around 12.5 nm in the spectrum) to achieve better contrast. The band center locations are defined by the location where the differential is equal to 0.

**ASSOCIATED CONTENT**

**Supporting information**

Supplemental Video S1 for FF-tr and NF-tr of the single layer alloy TMD.

Supplemental Video S2 for NF-tr and NF-PL of the bilayer alloy TMD.

**ACKNOWLEDGMENT**

We thank Prof. Yia-Chung Chang, Dr. Ching-Hang Chien, and Prof. Wen-Hao Chang for insightful discussions. We acknowledge the focused ion beam facility support in the Research Center for Applied Sciences, Academia Sinica. This work is supported by the "Innovative Instrumentation" project (AS-CFII-108-203) of Academia Sinica and by the Ministry of Science and Technology of Taiwan under the grant number MOST 109-2112-M-001-038 and MOST 110-2112-M-001-059.